\documentclass[aps,prl,twocolumn,groupedaddress]{revtex4}

\usepackage{graphicx}
\usepackage{amssymb}

\newcommand{\rs}{\rm \scriptscriptstyle}

\begin{document}

\title {Microscopic derivation of Hubbard parameters for cold atomic gases}

\author{Hans Peter B\"uchler}
\affiliation{Institute for Theoretical Physics III, University of Stuttgart, Germany}

\date{\today}

\begin{abstract}
We study the exact solution for two atomic particles in an optical lattice interacting via
a Feshbach resonance. The analysis includes the influence of all higher bands, as well as 
the proper renormalization  of molecular energy in the closed channel.
Using an expansion in Bloch waves, we show that the problem reduces to a simple matrix equation,
which can be  solved numerically very efficient. This exact solution allows for the precise determination 
of the parameters in the Hubbard model and the two-particle bound state energy. We identify the regime, where
a single band Hubbard model fails to describe the scattering of the atoms as well as the bound states.
%
\end{abstract}


\maketitle

Cold atomic gases in optical lattices represent a perfect laboratory system 
for the quantum simulation of strongly correlated many-body systems described 
by Hubbard models \cite{greiner02,jaksch04,bloch08}.  Recently, experimental and theoretical efforts focus on 
the  observation of a Fermionic Mott insulator \cite{jordens09,schneider08}, and the ultimate goal towards
the realization of magnetic and superconducting phases. The quantitative 
understanding of these experimental results and the comparison with the theoretical
predictions require a precise knowledge of the parameters in the Hubbard model
for cold atomic gases interacting with a Feshbach resonance \cite{leo08}. In this letter, we present 
the solution to the two-particle problem in an optical lattice 
interacting via a Feshbach resonance, and provide a microscopic derivation of the
parameters in the Hubbard model and the two-particle bound state energies.

\begin{figure}[ht]
 \includegraphics[width= 0.9\columnwidth]{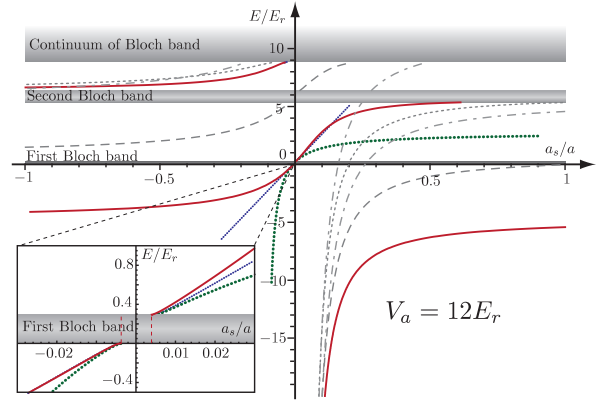}
  \caption{Exact bound state energies (red line) for two-atoms in an optical lattice of strength $V=12 E_{r}$ and ${\bf K}=0$. The additional
  bound states (dashed grey lines)  are weakly 
  coupled to atoms in the lowest Bloch band, i.e., $|w_{\alpha}|^2 \lesssim 10^{-4}$.  The green dotted line denotes the
  bound state energies predicted from the Hubbard model with the on-site interaction determined by Eq.~(\ref{onsiteinteraction});
  its deviations from the exact bound state indicate the break down of the Hubbard model. The blue dotted line denotes
  the bound states neglecting the correction $\chi_{0}-|w_{0}|^2 G(0)=-3.5$.}   \label{fig1}
\end{figure}

The two-particle interaction potential between particles in ultra-cold atomic gases is well described
by the pseudo-potential or in the presence of a Feshbach resonances within a two-channel 
model \cite{petrov05,bloch08}. The two-particle problem within confined geometries has extensively 
been studied for the one-dimensional setup with strong transverse confining \cite{olshanii98,bolda03}, and
the harmonic trapping potential \cite{busch98}.  In addition, the influence of optical lattices has been studied
for the deep lattices, where  the influence of higher bands have been included semiclassicaly \cite{fedichev04}
or using the exact solution for a harmonic oscillator within each well of the lattice  \cite{dickerscheid05,diener06,wouters06}.
%
%
Here, we analyze the two-particle problem interacting via a Feshbach resonance 
in a three-dimensional optical lattice and show that the equations can be efficiently solved 
numerically. The solution provides the exact scattering properties and bound
state energies of two-particles in an optical lattice of arbitrary strength, see Fig.~\ref{fig1}.

From the exact scattering amplitude, we find the microscopic derivation for 
the interaction parameters in the Hubbard model. 
The simplest Hubbard model describes bosonic particles with creation (annihilation) 
operators $b_{i}^{\dag}$ ($b_{i}$),  and on-site interaction $U$
(extension to fermionic particles with spin is straightforward),
\begin{equation}
	H = -  \sum_{ i,j} t_{i j} b^{\dag}_{ i} b_{ j}  + 
	 \frac{U}{2} \sum_{i}  b_{i}^{\dag}b^{\dag}_{i} b_{i} b_{i} .
	 \label{hubbardmodel}
\end{equation}
The hopping energies $t_{i j}$ derive from a single particle band structure calculation,
and are related to the dispersion relation in the lowest Bloch band 
$E_{0}^{a}({\bf q}) = - 2 \sum_{i} t_{i j} \cos {\bf q}\left( {\bf R}_{i}-{\bf R}_{j}\right)$.
In turn, the on-site interaction $U$ is conventionally derived for weak interaction strengths
and deep optical lattices by replacing the exact pseudo-potential by a $\delta$-function interaction 
and restricting the system to the lowest Bloch band  \cite{jaksch98}; the latter step corresponds to introducing a
short distance cut-off  $\Lambda$ comparable to the lattice spacing $a$. This approach
is restricted to weak interactions with $|a_{s}|/a \ll 1$; here,  $a_{s}$ is the $s$-wave scattering length.
In the general situation, the precise derivation of the interaction potential in the Hubbard model for arbitrary interaction is obtained by 
comparing the exact scattering properties of two particles in an optical lattice with the scattering amplitude 
predicted from the Hubbard model. This approach is in analogy to the description of the interaction 
bewteen cold gases in free space in terms of a pseudo-potential:  the strength of the pseudo-potential is fixed
by the condition to reproduce the exact scattering properties.

In the following, the interaction between the two-particles is given by a 
Feshbach resonance, which can be conveniently described by the two-channel
approach. Then, the Feshbach resonance is characterized by the detuning $\nu$ and 
the coupling $g$ between the open and closed channel, and gives rise to the 
scattering amplitude \cite{petrov05}
\begin{equation}
 f({\bf k}) =- \frac{2\mu}{4 \pi \hbar^2} \frac{g^2}{\epsilon_{{\bf k}}- \nu +
\frac{\mu g^2}{2 \pi \hbar^2} i \:k } \equiv -\frac{1}{ \frac{1}{a_{s}} + i k + r_{b}
k^2} \end{equation}
with ${\bf k}$ the incoming momentum and $\mu$ the reduced mass. The scattering 
length $ a_{s}=  -2 \mu g^2/4 \pi \hbar^2 \nu$ and the effective range  
$r_{\rs b} =   \pi \hbar^4/\mu^2 g^2$  are experimentally
accessible by measuring the bound state energy of the molecules across
the Feshbach resonance \cite{claussen03}.  

The two particles in the open
channel are described by the wave function $\psi({\bf x},{\bf y})$ with ${\bf
x}$ and ${\bf y}$ the position of the particles. In order to capture the above
characteristics of a Feshbach resonance, it is enough to describe the closed
channel by a single molecular state $\phi({\bf z})$. Then, the Schr\"odinger equation for
the energy eigenstates reduces to
\begin{eqnarray}
  \left[ E -H_{0}^{a}+H_{0}^{b} \right]\psi({\bf x},{\bf y})&\!= \!&
 g \int d{\bf z} \alpha({\bf r}) \phi({\bf z}) \delta\left({\bf z}-{\bf R}\right)   \label{twochannelmodel}
\\ \nonumber
  \left[E -  \nu_{0} - H_{0}^{m} \right]  \phi({\bf z})& \!=\! & g \int d{\bf x} d{\bf y} \alpha({\bf r})  
 \psi({\bf x},{\bf y}) \delta({\bf z}\! -\!{\bf R}),
\end{eqnarray}
where the single particle physics is described by the Hamiltonians
$H_{0}^{\sigma} = -\frac{\hbar^2}{2 m_{\sigma}} \Delta + V_{\sigma}({\bf x}) $
with $V_{\sigma}$ the optical lattices and the molecular mass $m_{m} = m_{a}+m_{b}$. 
Furthermore, we have introduced the relative 
${\bf r} = {\bf x}-{\bf y}$ and center of mass coordinates
${\bf R} = (m_{a}{\bf x}+m_{b}{\bf y})/(m_{a}+m_{b})$.
The properties of the Feshbach resonance are determined by the coupling strength 
$g$ and the bare detuning $\nu_{0}$, while  $\alpha({\bf r})= \exp(- {\bf r}^2/2 \Lambda^2)/ (2 \Lambda^2 \pi)^{ 3/2}\rightarrow \delta({\bf r})$
accounts for a regularization of the coupling with cut-off $\Lambda$. 
In the limit $\Lambda \rightarrow 0$, the bare detuning  $\nu_{0}$ entering the microscopic theory  is related to the physical observable detuning $\nu$
via $\nu_{0} = \nu - \nu_{\rs ren}$ with the renormalization $\nu_{\rs ren}=- g^2 \mu/(2 \hbar^2 \pi^{3/2} \Lambda)$.

The periodic structure of the optical lattice is characterized by the lattice vectors $\{{\bf R}_{j}\}$.
The single particle properties are then fully determined by the Bloch
wave functions $ \psi^{a}_{n,{\bf k}_{a}}({\bf x}) $, $\psi^{b}_{m,{\bf q}}({\bf y})$, and 
$\phi_{s,{\bf K}}({\bf z})$ with the corresponding band energies 
$E^{a}_{n}({\bf k}_{a})$, $E^{b}_{n}({\bf k}_{b})$, and $E^{m}_{s}({\bf K})$; 
here, ${\bf k}_{a}$, ${\bf k}_{b}$ and ${\bf K}$ are 
the quasi-momentum,  while $n$, $m$, and $s$ characterize the different
Bloch bands. In the following, we measure energies with
respect to the ground state energy of two-particles in the lowest Bloch band, i.e.,
$E^{a}_{0}(0)+E^{b}_{0}(0)=0$. The discrete translation invariance provides  
the conservation of the total quasi-momentum ${\bf K} = {\bf k}_{a}+{\bf k}_{b}$. 
Then, the general solution with fixed total quasi-momentum ${\bf K}$ can 
be written as 
\begin{displaymath}
  \psi({\bf x},{\bf y}) =  \frac{1}{\sqrt{N}}\sum_{n,m} \sum_{{\bf q}} \varphi^{n m}({\bf q})
\psi^{a}_{n,{\bf q}}({\bf x}) \psi^{b}_{m,{\bf K}-{\bf q}}({\bf y}), 
\end{displaymath}
and $\phi({\bf z}) =  \sum_{s} R^{s} \phi_{s,{\bf K}}({\bf z}) $.
Inserting this expansion in Eq.~(\ref{twochannelmodel}), we obtain 
\begin{eqnarray}
  \left[E - E_{n m}({\bf q})\right] \varphi^{n m}({\bf q}) & = &
w \sum_{s} h^{n m}_{s}({\bf q}) R^{s}  \label{latticepsiequation}\\ 
\left[E- \nu_{0}- E_{s}^{m}\left({\bf K}\right)\right]R^{s}
& = & w \sum_{n, m} \frac{1}{N} \sum_{{\bf q}} h^{s}_{n m}({\bf q})  \varphi^{n m}\left({\bf
q}\right). \nonumber
\end{eqnarray}
Here, we have introduced the notation $E_{n m}({\bf q})= E^{a}_{n}({\bf q}) + E^{b}_{m}({\bf K}-{\bf q})$, and
the characteristic coupling energy $w = g/\sqrt{V_{0}}$ with $V_{0}$ the volume of the unit cell.
The dimensionless coupling elements  reduce to
\begin{displaymath}
  \frac{ h^{nm}_{s}({\bf q})}{\sqrt{N V_{0}}}
   = \int \! d{\bf x} d{\bf y} \!\left[\psi^{a}_{n,{\bf q}}({\bf x}) \psi^{b}_{m, {\bf K}-{\bf q}}({\bf y})\right]^{*}\!\! \alpha({\bf r})
\phi_{s,{\bf K}} \left({\bf R}\right),
\end{displaymath}
with the notation $h^{s}_{n m}({\bf q}) = \left[h^{n m}_{s}({\bf q})\right]^{*}$, while $N V_{0}$ denotes the quantization volume.
Substituting the bare
detuning $\nu_{0}$ with the physical detuning $\nu$ by adding on both sides of Eq.~(\ref{latticepsiequation})
the renormalization $\nu_{\rs ren}$, we obtain
\begin{equation}
\left[E - \nu - E_{s}^{m} \right] R^{s}-
w^2 \sum_{t}\chi^{s}_{t}(E) R^{t}  = 0.
\label{requation} 
\end{equation}
The matrix $\chi^{s}_{t}$ describes the shift of the Feshbach resonance due to
the change in dispersion relation of the particles in the open channel; this phenomena is in analogy to
the lamb shift of atoms in a cavity \cite{brune94}. It takes the
form ($v_{0}$ denotes the volume of the Brioulline zone)
\begin{displaymath}
 \chi^{s}_{t}(E)\!= \!\sum_{n, m} \int \frac{d{\bf q}}{v_{0}} 
 \left[\frac{h_{n m}^{s}({\bf q}) h^{n m}_{t}({\bf q})}{E- E_{n m}({\bf q}) + i \eta} 
  + \frac{\hat{h}_{n m}^{s}({\bf q}) \hat{h}^{n m}_{t}({\bf q})}{\hat{E}_{n
m}({\bf q}) }\right]
\end{displaymath}
The quantities 
$\hat{E}_{n m}({\bf q})$ and $\hat{h}^{n m}_{s}({\bf q})$ are the energies
and coupling parameters for the system in absence of an optical lattice.
The first term in the above equation describes the influence of higher bands, while
the second term appears from the  renormalization. The divergent parts in the two terms 
cancel each other, and $\chi_{t}^{s}$ remains finite in the limit $\Lambda\rightarrow 0$. 
This behavior can be easily understood:
for large Bloch bands, the influence of the optical lattices vanishes and the coupling elements
$h_{s}^{nm}({\bf q})$ and energies $E_{n m}({\bf q})$ reduce to the values of the free system. Then, the
terms in the  bracket cancel each other, and the summation over the Bloch bands converges. For a finite short
distance cut-off $\Lambda$,  the corrections vanishes with $\sim \Lambda$; i.e., the convergence is very slow in the number of
Bloch bands.

In the following, we discuss the setup with a three-dimensional
cubic lattice  $V_{\sigma}({\bf x}) =V_{\sigma} \sum_{i=1}^{3} \sin^2( k_{\rs L} x_{i})$ 
with lattice spacing $a=\pi/k_{\rs L}$ and  recoil energy
$E_{r}= \hbar^2 k_{\rs L}^2/2 m$. For equal particle species $m_{a}=m_{b}$ 
and far detuned optical lattice, the relative strengths of the lattice potentials 
naturally satisfy $V_{a}=V_{b}=V_{m}/2$. We focus on a {\it wide} 
Feshbach resonance; the generalization to a {\it narrow} Feshbach 
resonance is straightforward.
A wide Feshbach resonance is obtained in the
limit $\nu, g \rightarrow \infty$ with a fixed $s$-wave scattering length 
$a_{s}=  -m g^2/4 \pi \hbar^2 \nu$, and the
energies $E$ and $E_{s}^{m}$ in the first term in Eq.~(\ref{requation}) can be dropped.

\begin{figure}[ht]
 \includegraphics[width= 1\columnwidth]{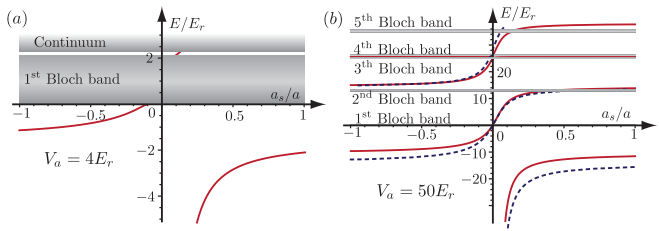}
  \caption{Bound state energies (red line) for different lattice depths: 
  (a)  For weak optical lattices with $V=4 E_{r}$, 
  the Bloch bands in three-dimensions nearly overlap. (b) For deep optical lattices with $V=50 E_{r}$, 
  the bound states energies are compared with the energies obtained by replacing the optical lattice by 
  a harmonic well with trapping frequency $\omega_{p}=\sqrt{4 V E_{r}}/\hbar$ (dashed blue line). }   \label{fig2}
\end{figure}

{\it Bound states}: The equation for the energies $E_{\rs B}$ of the repulsive and attractive bound states 
reduces to the eigenvalue equation
%
\begin{equation}
    \sum_{t} \left[ \delta ^{s}_{t}+ W \: \chi^{s}_{t}(E_{\rs B})\right] R^{t}=0
\end{equation}
with $R^{s}$ the molecular wave function of the bound state, and 
$W= - g^2/\nu V_{0} =  (8/\pi) E_{r} a_{s}/a$; 
the numerical solution is shown in Fig.~\ref{fig1} and Fig.~\ref{fig2}.
For a fixed value of the $s$-wave 
scattering length $a_{s}$ and quasi-momentum $K$, there  appear now several
bound states. This behavior is in strong contrast to the free system, where for a fixed 
center of mass motion only a single bound state exists for a repulsive scattering length $a_{s}$.
Here, the appearance of several bound states is a consequence of the reduced translation symmetry: 
molecular states and two-particles states in the open channel differing in center of mass motion by a reciprocal 
lattice vector are coupled the periodic lattice. These coupling strength are given by the overlaps 
$h^{n m}_{s}({\bf q})$. 

The matrix elements $\chi^{s}_{t}(E)$ are fully determined by the single particle 
properties such as the Bloch wave functions and Band structure, and can be efficiently 
determined numerically.  For the three-dimensional cubic lattice above, the single particle 
wave function separate for each space direction, i.e., 
$u_{n,{\bf k}}({\bf x})= \prod_{i=1}^{3} u^{\rs 1D}_{n_i,k_i}(x_{i})$,  and it is therefore 
sufficient to determine the Bloch wave functions and energies for a one-dimensinal setup;
the Bloch wave functions are determined using $Z=201^3$ reciprocal lattice vectors.
The integration and the summation over the different Bloch bands is performed. 
While the integration converges very quickly using $N= 41^3$ unit cells, the limiting factor in accuracy 
for the matrix elements $\chi^{s}_{t}(E)$ is the  slow convergence with the number of Bloch bands: 
the restriction to a finite number of Bloch bands in the summation corresponds to introducing a
high energy cut-off $\Lambda \sim 1/(k_{\rs L}S^{1/3})$. Consequently, the summation converges
with  $ \sim 1/S^{1/3}$, and a finite size scaling analysis can be performed; the number of Bloch bands included in
this analysis was $S=11^3$. Then, the matrix elements $\chi^{s}_{t}$ 
can be calculated with an accuracy better than $1\%$; the convergence has been extensively tested for
varying  number of unit cells $N=41^3 \ldots 201^3$, $S=11^3\ldots 21^3$, and $Z=100^3 \ldots 500^3$. 

{\it Scattering amplitude}: Next, we analyze the scattering states in the lowest Bloch band;
%
\begin{eqnarray}
\varphi^{n m}({\bf q}) &= & \varphi_{0}^{n m}({\bf q})   + 
 \frac{\lambda^{n m}({\bf q},{\bf k},{\bf K})}{E-E_{n m}({\bf q},{\bf
K})+ i \eta}
\end{eqnarray}
with $\varphi_{0}({\bf q}) = \delta_{n,0} \delta_{m,0} \delta_{{\bf q},{\bf k}}$ an incoming 
wave at relative momentum ${\bf q}$ and center of mass momentum ${\bf K}$ in the lowest Bloch band. Then the generalization 
of the $s$-wave scattering length in free space is obtained via  
$\lambda \equiv  \lambda^{0 0}({\bf q}\rightarrow 0,0,0)$ at low energy $E \rightarrow 0$ of the incoming wave.
Again, the scattering amplitude $\lambda$ is fully determined by the matrix $\chi^{t}_{s}$.
 Introducing the notation $R^{s}_{\alpha}$ for the eigenvectors of the 
matrix $\chi^{s}_{t}$ with eigenvalues $\chi_{\alpha}$, the scattering amplitude
reduces to ($W\equiv (8/\pi) E_{r} a_{s}/a$)
\begin{equation}
  \lambda  = W  \sum_{\alpha}  \frac{  |w_{\alpha}|^2}{1- W \chi_{\alpha}} \approx \frac{W |w_{0}|^2}{1- W \chi_{0}} 
\end{equation}
with $w_{\alpha} = \sum_{s} h^{00}_{s}(0) R^{s}_{\alpha}$ the width of each scattering resonance.
The crossing of each bound state with the lowest Bloch band, see Fig.~\ref{fig1}, gives rise to a pole in the scattering amplitude
and describes a scattering resonance. Except for the first resonance, the couplings are in general weak and 
the scattering amplitude is dominated by the lowest eigenvalue $\chi_{0}$
and width $w_{0}$. However, these additional resonances can give rise to characteristic loss features for cold atoms in an optical lattice
at large $s$-wave interactions.

In the following, we will now compare this exact value for the scattering amplitude for two particles in an
optical lattice with the predictions from the Hubbard  model Eq.~(\ref{hubbardmodel}). For an on-site interaction 
$U$ the scattering solution in the Hubbard model takes the form  \cite{winkler06}
\begin{equation}
\varphi_{\rs HM}({\bf q}) = \delta_{{\bf q},{\bf k}} +  \frac{\lambda_{\rs HM}}{E-E_{0 0}({\bf q},{\bf K})+ i \eta},
\end{equation}
where $\lambda_{\rs HM}$ describes the scattering amplitude in the Hubbard model;
 $ \lambda_{\rs HM} = U/[1- U G(E)]$ with $G(E) = \int \frac{d{\bf q}}{v_{0}} [E\!-\!E_{00}({\bf q})\!+\!i\eta]^{-1}$. 
For nearest neighbor hopping $t$ and low scattering energies $E\rightarrow 0$,  
$G(0)$ reduces to $G(0)= c/2 t$ with $c\approx -0.2527$.
The effective on-site interaction $U$ is therefore completely fixed by the condition,
that the Hubbard model reproduces the exact two-particle properties, i.e., $ \lambda_{\rs HM} \equiv \lambda$,
and we obtain
\begin{equation}
  U = \frac{1}{ \lambda^{-1}+G(0)} \approx  \frac {W |w_{0}|^2}{1- W \left[\chi_{0}- |w_{0}|^2 G(0)\right]} .
  \label{onsiteinteraction}
\end{equation}
The parameters for different strengths of the optical lattice are shown in Table~\ref{table1}. 
The contribution $W |w_{0}|^2$ describes the dominant part for weak interactions,  while the correction
$ \chi_{0}- |w_{0}|^2 G(0)$ becomes relevant for stronger interactions.
It is important to stress, that this derivation of the on-site interaction $U$ is valid for arbitrary values of the
$s$-wave scattering length $a_{s}$, and gives rise to a finite value $U_{\infty}$ for $a_{s}\rightarrow \pm \infty$.
However, its validity is restricted to low scattering energies: first, additional interaction terms beyond the on-site interaction
$U$ can play an important role and will account for the full momentum dependence of the scattering amplitude
$\lambda^{n m}({\bf q},{\bf k},{\bf K})$. Second, the bound state energies can be strongly modified by additional terms, which are 
not included in a single band Hubbard model. A test for the validity of the Hubbard model is therefore the
comparison with the exact bound state energy and the repulsive/attractive bound states predicted from the
Hubbard model. The bound states within the Hubbard model are determined by 
poles in the scattering amplitude $\lambda_{\rs HM}$, i.e., $U G(E)=1$. A comparison with the
exact bound state energies is shown in Fig.~\ref{fig1}, and we find already  very strong deviations 
for $a_{s}/a\gtrsim 0.02$ at $V=12 E_{r}$: the validity for the description of bound states in the
Hubbard model is limited to very weak interactions.

\begin{table}
\begin{tabular}{c| c c c c }
\hspace{5pt} $V/E_{r}$ \hspace{5pt} &  \hspace{10pt}  $t/E_{r}$  \hspace{10pt}    &\hspace{10pt}  
 $\chi_{0} E_{r}$  \hspace{5pt}  &  \hspace{5pt}  $|w_{0}|^2$  \hspace{5pt}  & $\chi_{0}\!-\! |w_{0}|^2 G(0) $\\
 \hline 
 4 &0.0855  &-4.188 &2.412 &$-0.6/E_{r}$\\
 8 &0.0308  &-26.82&5.954 & $-2.3/E_{r}$\\
 12 &  0.0122  & -101.3 & 9.483 &$ -3.5 /E_{r} $\\
  16 & 0.00533 &  -303.0 &12.63 & $-3.7/E_{r}$\\
 20 & 0.00249 & -788.2 &15.50& $-1.8 /E_{r}$\\
\end{tabular}
\caption{Effective parameter in the Hubbard model with hopping $t$. The on-site interaction $U$ is given by Eq.~(\ref{onsiteinteraction}).  \label{table1}}
\end{table}

Finally, for  {\it deep optical lattices} $V/E_{r}> 1$, the width of the lowest Bloch 
band provides a small parameter characterized by the the hopping energy $t/E_{r}\ll 1$. As a consequence, 
for all energies  $E$ of the order of the band width $E\sim 12 t$,  the first term in the 
matrix $\chi_{t}^{s}$ dominates
\begin{displaymath}
 \chi^{s}_{t}(E) \approx \int \frac{d{\bf q}}{v_{0}} \frac{h_{0 0}^{s}({\bf q}) h^{0 0}_{t}({\bf q})}{E- E_{0 0}({\bf q}) + i \eta}  \sim \frac{1}{ 12 t},
\end{displaymath}
while all the remaining terms from higher Bloch bands as well as the renormalization provide a contribution $\sim 1/E_{r}$. 
Consequently, the results reduces to the well known approach \cite{jaksch98} for the derivation of the Hubbard parameters,
where the influence of higher bands are neglected and the pseudo potential is replaced by a $\delta$-function. 
Then, the momentum dependence
of the interaction potential in the Hubbard model reduces to

\begin{displaymath}
    U({\bf q},{\bf k},{\bf K})   =  \frac{g^2}{
\nu}\int d{\bf z} \left[\psi^{a}_{0,{\bf q}}\psi^{b}_{0, {\bf K- q}}\right]^{*}\psi^{a}_{0,{\bf k}} \psi^{b}_{0, {\bf
K-k}}.
\end{displaymath}
This term also accounts for contributions such as nearest-neighbor interactions and correlated hopping \cite{jaksch98,duan05,werner05}.
For increasing interactions, the shift $\chi_{0}\!-\! |w_{0}|^2 G(0)\sim 1/E_{r}$ in Eq.~(\ref{onsiteinteraction}) becomes important, 
and in addition the bound state energies $E_{\rs B}$ start to deviate from the predictions within the Hubbard model.
The crossover from the
two regimes can be self-consistently checked: the higher bands become relevant as soon as $E_{\rs B}$ becomes in the range
of the Bloch band separation $U\sim \hbar \omega_{p}$ , while the renormalization requires
$W \ll E_{r}$, i.e., $a_{s}\ll a$. In the limit of deep optical lattices, the first condition is always more stringent  and reduces
to $a_{s}/a\ll (E_{r}/V)^{1/4}/2 \sqrt{\pi} $, which has previously been suggested \cite{jaksch98}. In order to derive Hubbard models which
reproduce the bound state as well as the scattering states, the influence of the higher bands as well as the renormalization have to be included;
in contrast to recent attempts to include the influence of higher bands alone \cite{duan05}.


We would like to thanks  L. Tarruel, T. Esslinger, M. Troyer, P. Zoller, and A. Muramatsu for fruitful discussions. Support from the DFG within
SFB/TRR21 and DRAPA OLE program is acknowledged.


\end{document}